\documentclass{Interspeech2024}
\usepackage{hyperref, url}
\usepackage[utf8]{inputenc} 
\usepackage[T1]{fontenc}

\usepackage{multirow}
\usepackage{adjustbox}
\usepackage{colortbl}
\usepackage{highlight}

\interspeechcameraready

\title{HLTCOE JHU Submission to the Voice Privacy Challenge 2024}

\name[]{Henry Li}{Xinyuan}
\name[]{Zexin}{Cai}
\name[]{Ashi}{Garg}
\name[]{Kevin}{Duh}
\name[]{Leibny Paola}{Garc\'ia-Perera}
\name[]{Sanjeev}{Khudanpur}
\name[]{Nicholas}{Andrews}
\name[]{Matthew}{Wiesner}

\address{
  Johns Hopkins University, United States} 
\email{xli257@jhu.edu}


\begin{document}

\maketitle

\begin{abstract}
    We present a number of systems for the Voice Privacy Challenge, including voice conversion based systems such as the kNN-VC method and the WavLM voice Conversion method, and text-to-speech (TTS) based systems including Whisper-VITS. We found that while voice conversion systems better preserve emotional content, they struggle to conceal speaker identity in semi-white-box attack scenarios; conversely, TTS methods perform better at anonymization and worse at emotion preservation. Finally, we propose a random admixture system which seeks to balance out the strengths and weaknesses of the two category of systems, achieving a strong EER of over $40 \%$ while maintaining UAR at a respectable $47\%$. 
    

\end{abstract}

\section{Introduction}

Technology has enabled unprecedented means for extracting personally identifiable information from speech. New legislation that recognizes this reality including the Illinois biometric privacy act \cite{illinois_privacy}, California's consumer privacy act of 2018, South Africa's Protection of personal information act (POPIA), and most notably, the EU's General Data Protection Regularion (GDPR), greatly limit the extent to which personally identifiable information, including speech can be stored or processed. Compliance with these laws, and protection of users' private information requires means of anonymizing a user's voice without corrupting the data to such an extent that it no longer has any linguistic or paralinguistic content. The Voice Privacy Challenge is on initiative driving these efforts.

In the Voice Privacy Challenge voice anonymization is modeled as a user-attacker game.
A \textit{user} anonymizes their speech with the goal of preserving linguistic and emotional content, while an \emph{attacker} attempts to identify speech from any remaining signal in the anonymized utterances including the remaining linguistic and emotional content. The challenge specifies that the Voice Anonymization system is a function $f$ that operates on the utterance level. 

To measure these two competing objectives, challenge submissions are evaluated using utility and privacy metrics. Utility refers to the extent to which the spoken and emotional is preserved following anonymization. These metrics are measured using a pre-trained systems: we note that this approach makes it possible to optimize directly on these metrics. Content preservation is measured by Word Error Rate (WER) in a traditional Automatic Speech Recognition (ASR) task setup. Emotion is measured by Unweighted (across the different emotion labels) Average Recall (UAR). Privacy is measured using an ECAPA-TDNN-based \cite{ecapa_tdnn} Automatic Speaker Verification (ASV) system. At training time, a classifier is attached to the ECAPA-TDNN model and trained jointly on speaker labels; at inference time, ECAPA-TDNN representations are extracted for a set of ``enrollment utterances" --- utterances known to have been produced by a certain speaker, and for ``test utterances" - utterances whose speaker the attacker endeavors to determine. A distance metric (cosine distance) is computed for each test utterance relative to the pool of enrollment utterances. A distance threshold, where accuracy equals recall if all utterances whose distance to the enrollment pool is less than the threshold are taken to belong to enrollment speaker, is identified. The resulting error rate is known as Equal Error Rate (EER). The theoretical upper limit of EER is $50\%$.

Anonymization systems tend to excel at either utility preservation or anonymization, but not both. To address this short-coming, we propose two different anonymization methods that excel on each individual task and a simple, but effective method to combine these approaches that enables a reasonable tradeoff between the competing objectives. First we describe a simple voice-conversion approach that focuses on preserving linguistic and emotional content, but not anonymization under the strong attack model in the Voice Privacy Challenge. Inspired by the anonymization approach by Meyer et al.~\cite{meyer2023prosody}, we also adopt a cascaded ASR-TTS approach for anonymization. However, the original approach replicates the prosodic features of the original speech, which can lead to speaker identity leakage in this year’s challenge, where attackers have access to anonymized speech for training speaker verification models. To mitigate this, we use a multi-speaker TTS system to eliminate the speaker footprint. This cascaded approach excels in anonymization but degrades the utility of the speech signal. Finally, we describe our method for combining both approaches.

\section{Method}

\begin{figure*}[ht]
  \centering
  \includegraphics[width=0.9\textwidth]{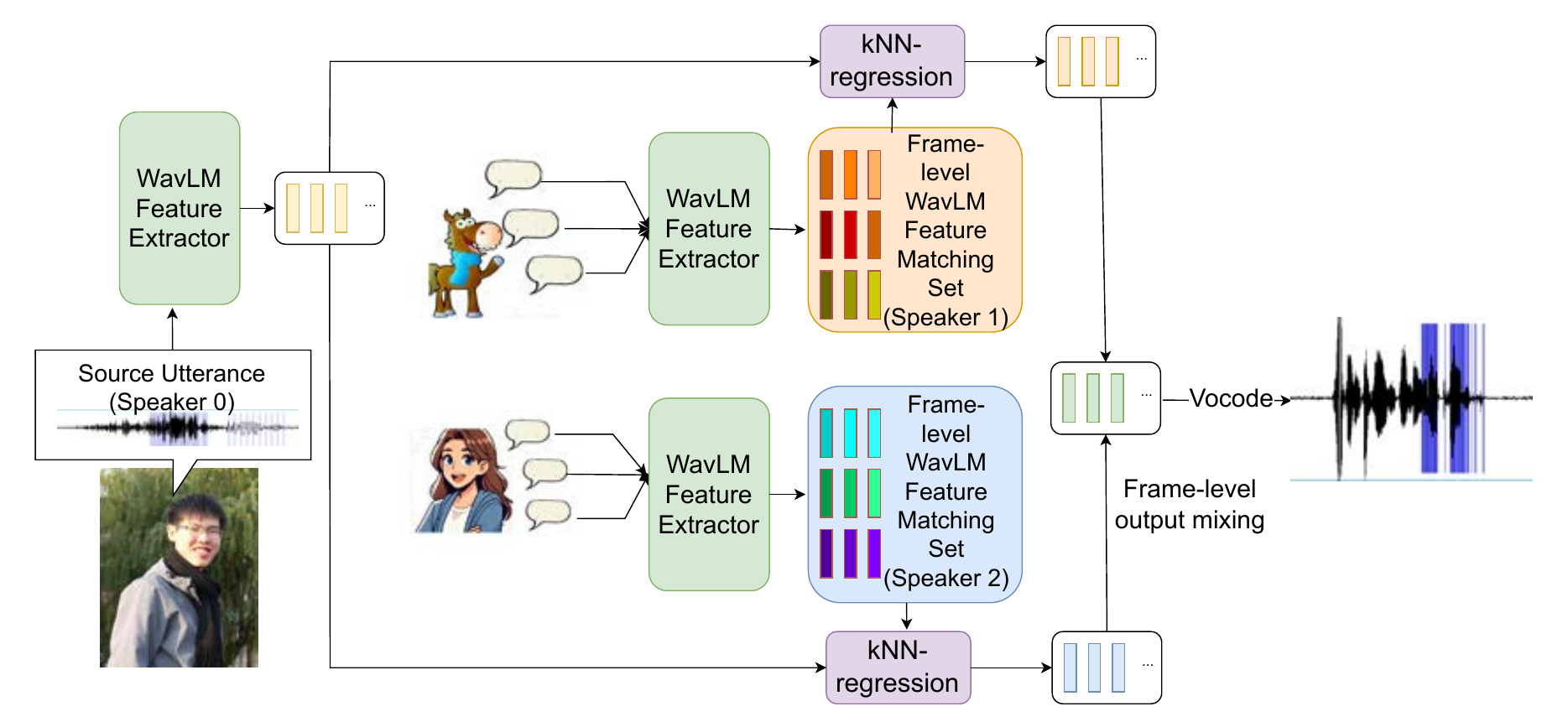}
  \caption{Schematic of our adapted kNN-VC system.}
  \label{fig:knn}
\end{figure*}

\subsection{Our Baseline: kNN-VC}

Our baseline system extends earlier work by Baas et al. \cite{knn_vc}. The method, as with our subsequent adaptations, operate at WavLM-feature level \cite{wavlm}. The original kNN-VC system consists of a simple yet elegant idea: first, both the source utterance and a number of utterances from the target speaker (totalling at least $5$ minutes in length) are converted into the WavLM feature space; next, k-nearest neighbour regression is performed on each frame in the WavLM-feature representation of an utterance with respect to the set of all WavLM feature frames in the target speaker's utterance pool. The average of the k-nearest neighbour output is used to synthesize the target utterance using a HiFi-GAN vocoder \cite{hifigan} trained to synthesize speech using WavLM feature vectors. The vocoder we used in our system was trained on the LibriSpeech train-clean-100 subset, matching the one used by the original authors of kNN-VC. We picked $k = 4$ during our k-nearest neighbour regression, matching the setup in the original paper.

The following modifications to the original kNN-VC system (ID = 1 in  Table \ref{tab:results}) were shown to perform marginally superior at anonymization: choosing random target speakers; applying length variation (ID = 2); applying an additive noise to the WavLM features pre-conversion (ID = 3). These improvements come at the cost of emotion or content preservation.

The original kNN-VC system targets the voice characteristics of a single speaker, which may run contrary to the overarching goals of privacy protection. In order to address this problem, we revised the kNN-VC system so that an arbitrary number of matching sets, each originating from a different speaker/pseudo-speaker, is prepared. At inference time, the kNN outputs from each matching set is pooled together using a weighted average. The schematic of our revised kNN-VC system is shown in figure \ref{fig:knn}.

\begin{figure*}[ht]
  \centering
  \includegraphics[width=0.9\textwidth]{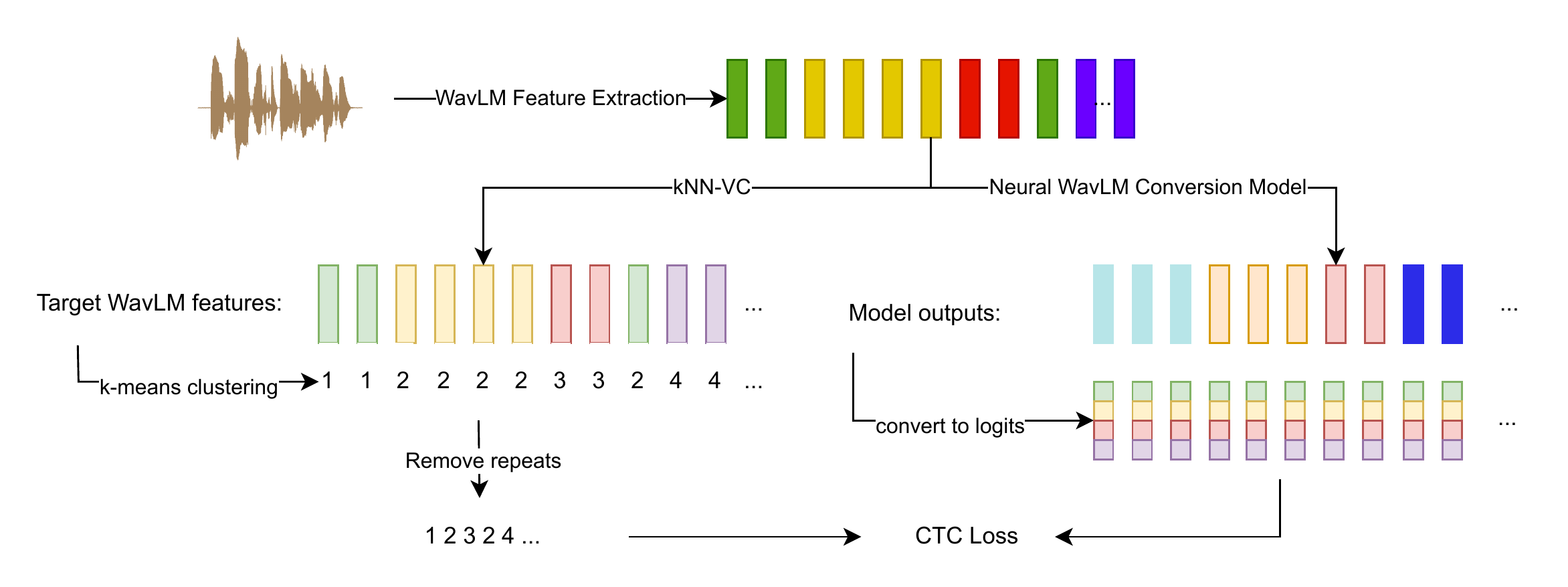}
  \caption{Schematic of our WavLM conversion system with k-means discrete loss. The training targets (on the left hand side) are discretized using k-means clustering. The resulting token sequence is used as the golden target labels during CTC loss calculation.}
  \label{fig:k_means}
\end{figure*}

\subsection{WavLM Conversion}

The kNN-VC system produces converted utterances that, qualitatively, have remarkably similar vocal characteristics to those produced by the target speaker. It is strong at preserving both the content and the emotion of the original utterance, outperforming all of the previous submissions. We do, however, note the relative weakness of the kNN-VC system on the privacy objective compared to cascaded TTS-based systems (such as \cite{asrbn}). The simplicity of this method also hinders any efforts to address this weakness. As such, we propose a transformer-based WavLM feature conversion model, in the hopes that a neural method would offer more flexibility to allow us to tackle this weakness.

One obstacle to training voice conversion models is the lack of parallel data from different speakers. Fortunately, our baseline kNN-VC system provides us with a virtually unlimited synthetic pool of such training data. As such, we propose the following definitions:

\begin{enumerate}
    \item $w$: WavLM feature extractor
    \item $k$: kNN-VC feature converter. $k(w(u), a)$ converts the WavLM features of utterance $u$ into that of target speaker $a$ using the kNN-VC system.
\end{enumerate}

Given these definitions, we proceed to train our WavLM feature conversion system to predict $k(w(u), a)$ from $w(u)$.

The base model architecture was adapted from FastSpeech2 \cite{fastspeech2}, a non auto-regressive, encoder-decoder TTS system. We adapted an implementation of FastSpeech2 from Chien et al. \cite{fastspeech2_code}, modifying the model task from TTS to Voice Conversion and changing the representation that the model operates on from Mel Spectrogram to WavLM features. Our model uses $4$ encoder layers and $6$ decoder layers, each with a hidden dimension of $512$.

In its most basic form, the model learns to faithfully replicate the output and therefore inheriting the weakness at anonymization of the kNN-VC system. Changes to the training procedure are necessary so as to improve in that regard.

Some of the additions we made to the base WavLM conversion model include:

\begin{enumerate}
    \item Joint training with target reconstruction objective (such that some $u$ belong to the target speaker ($u_a$))
    \item Joint Adversarial training with speaker ID task. A speaker-ID system is jointly trained with the WavLM conversion model. The WavLM conversion model is optimized to hinder the speaker-ID system from producing the correct prediction.
    \item Discretized and aligned objective
\end{enumerate}

We will proceed to describe each of these additions in details:

\subsubsection{Joint training with target reconstruction objective}

In every batch, a percentage of ($w(u)$, $k(w(u))$) pairs are replaced by ($w(u_a)$, $w(u_a)$), where $u_a$ denotes an actual utterance by the target speaker. We hope that, by inducing the model to learn to reconstruct large chunks of the target speaker's speech, it would pick up on sequence-level characteristics of the target speaker's speech patterns and reproduce them at inference time during voice conversion.

\subsubsection{Joint adversarial training}

An ECAPA-TDNN-based Speaker-ID model, similar to the one that is pre-trained prior to being used for ASV evaluation in the challenge, can be trained to work with WavLM features as well, achieving up to $99.6\%$ accuracy on the train-clean-360 split of LibriSpeech \cite{librispeech}. We leverage this fact to perform Joint Adversarial Training between voice conversion and speaker ID. Specifically, the speaker ID system and the voice conversion system are trained jointly. For each batch of samples, the speaker ID system gets its weights updated so that it learns to better classify the source speakers using the output of the voice conversion system; the voice conversion system, in addition to learning to voice convert, also learns to prevent the speaker ID system from being able to correctly classify the source speaker. 

\subsubsection{Discretized and aligned objective}

The original FastSpeech2 implementation used a combination of L1 and L2 losses against the training target - utterances converted using the kNN-VC system. As the loss operates on the frame-level, it encourages the model to preserve the sequence-level information of the source utterance such as speed or accent: information that leads to strong emotion preservation at the expense of weaker anonymization. In order to allow the model to warp an utterance in the time domain, we design the following training objective: first, we discretize the set of target speaker WavLM feature vectors using k-means clustering. Using the resulting k-means model, we discretize each kNN-VC converted source utterance into sequences of discrete tokens. We then collapse any repeated tokens in each discrete label sequence. At training time, cosine distance between each frame that the WavLM conversion model generates and each k-means centroid is used as a logit value, which is in turn used to compute a likelihood vector for each centroid at each frame. A CTC-loss \cite{ctc} is then computed using the frame-level likelihood vectors against the collapsed target-side discrete unit sequence. Figure \ref{fig:k_means} illustrates our WavLM convserion system with the discretized objective in action.

\subsection{Cascaded Anonymization}
Although voice conversion systems can effectively alter the acoustic characteristics related to the speech production organs of source speakers, the prosodic characteristics, which reflect their acquired speaking habits and styles, remain unchanged. We hypothesize that such prosodic characteristics are beneficial for preserving emotions, while also serving as patterns to identify the corresponding speaker. To address this, we cascade automatic speech recognition (ASR) and text-to-speech (TTS) models to enhance anonymization performance by altering the speaking style of the source speech.

\begin{figure}[ht]
  \centering
  \includegraphics[width=0.32\textwidth]{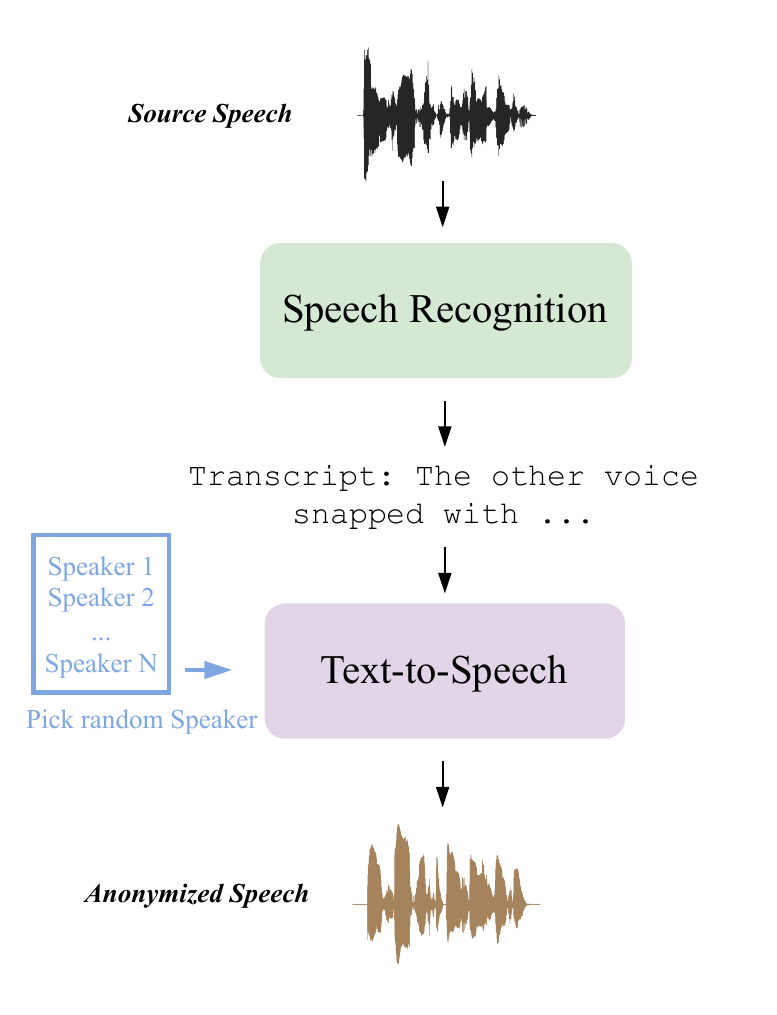}
  \caption{Cascaded ASR-TTS Anonymization Process}
  \label{fig:anon}
\end{figure}

The anonymization process of our cascaded system is illustrated in Figure~\ref{fig:anon}. We first obtain the transcript of the source utterance using an open-sourced ASR system. Then, we apply a TTS system to generate the corresponding anonymized utterance with a randomized voice. Specifically, we used the ‘medium-en’ model from Whisper\footnote{\url{https://github.com/openai/whisper}}~\cite{radford2023robust} as our ASR system to obtain the transcript. The synthesis system is an open-source multi-speaker TTS model\footnote{\url{https://huggingface.co/datasets/rhasspy/piper-checkpoints/blob/main/en/en_US/libritts_r/medium}}, VITS~\cite{kim2021conditional}, trained on the LibriTTS dataset. For anonymization purposes, we randomly selected a speaker’s voice from a pool of 904 speakers to synthesize each utterance. 

\subsection{Random Admixture}

As discussed earlier, our different systems exhibited different strengths and weaknesses: some excel at anonymization while others appear to do better at emotion preservation. Drawing inspiration from data poisoning attacks against neural networks \cite{poisoning}, which found that inserting poor quality or incorrect data into the training set can have a disproportional impact on model performance, we experiment with randomly selecting from one of our many anonymization systems during training at test time. We highlight the random admixture system where the two source pools we drew from are the Whisper-VITS TTS system and the original kNN-VC system. We note that while the emotion preservation performance is a linear extrapolation of that of the two source systems, the trained ASV system performed worse than the linear extrapolation of the EER of these two systems, as studies on data poisoning attacks would predict.  

\section{Experiments and Results}
\subsection{Datasets}
Subsets from the LibriSpeech~\cite{librispeech} and IEMOCAP~\cite{busso2008iemocap} corpora are used as development and evaluation data in the challenge. Detailed information can be found in the data description section of the challenge’s evaluation plan~\cite{tomashenko2024voiceprivacy}. Specifically, there are 10 subsets for the evaluation process. The subsets libri-dev-asr and libri-test-asr are used for ASR evaluation. The subsets libri-dev-enrolls, libri-dev-trials-f, libri-dev-trials-m, libri-test-enrolls, libri-test-trials-f, and libri-test-trials-m are used for evaluating anonymization (speaker verification) performance. The subset libri-train-clean-360 is used for training the speaker verification system after anonymization. For emotion preservation performance, IEMOCAP-dev and IEMOCAP-test are used. For our voice conversion systems, including our submitted kNN-VC system, we drew our target speaker(s) from the VoxCeleb \cite{voxceleb} dataset.

\subsection{Evaluation Metrics}
\subsubsection{Privacy}
The main metric used for privacy evaluation is the equal error rate (EER). EER is calculated based on similarity scores from each pair of utterances in the evaluation set. Such a pair is also known as a trial. The threshold $\theta$ is denoted as the decision boundary between same-speaker and different-speaker, while $P_{fa}(\theta)$ and $P_{miss}(\theta)$ represent the false alarm and miss rates at threshold $\theta$, respectively. The EER corresponds to the threshold $\theta_{EER}$ where $P_{fa}(\theta_{EER}) = P_{miss}(\theta_{EER})$. A lower EER signifies a higher speaker re-identification risk. Therefore, in the context of voice privacy, a higher EER indicates better privacy.

\begin{table*}[!th]
  \caption{Privacy and Utility Performance of Various Anonymization Approaches \\ (Darker Color Indicates Better Performance)}
  \label{tab:results}
  \adjustbox{max width=\textwidth}{
  \renewcommand{\arraystretch}{1.1}
  \centering
  \setlength{\tabcolsep}{3pt}
  \begin{tabular}{ll ccccc ccc ccc}
    \toprule
    \multirow{2}{*}{\textbf{ID}} & \multirow{2}{*}{\textbf{System}} & \multicolumn{5}{c}{\textbf{Privacy - EER} (\%) $\uparrow$} & \multicolumn{3}{c}{\textbf{Utility - UAR mean} (\%) $\uparrow$} & \multicolumn{3}{c}{\textbf{Utility - WER} (\%) $\downarrow$} \\ 
    \cmidrule(lr){3-7} \cmidrule(lr){8-10} \cmidrule(lr){11-13} 
    && libri-dev-f & libri-dev-m & libri-test-f & libri-test-m & avg. & IEMOCAP-dev & IEMOCAP-test & avg. & libri-dev & libri-test & avg. \\
    \midrule
    0 &  origin & \gradientcell{10.511}{0}{50}{white}{violet}{30} & \gradientcell{0.931}{0}{50}{white}{violet}{30}  & \gradientcell{8.761}{0}{30}{white}{violet}{30}  & \gradientcell{0.418}{0}{30}{white}{violet}{30}  &  \gradientcell{5.16}{0}{30}{white}{violet}{30}  & \gradientcell{69.0796}{0}{75}{white}{violet}{30}   & \gradientcell{71.0618}{0}{75}{white}{violet}{30}  &\gradientcell{70.07}{0}{75}{white}{violet}{30} & \gradientcell{1.807}{0}{5}{violet}{white}{30}  & \gradientcell{1.844}{0}{5}{violet}{white}{30} & \gradientcell{1.83}{0}{5}{violet}{white}{30}  \\
    1$^{\star}$ & kNN-VC & \gradientcell{11.789}{0}{50}{white}{violet}{30} & \gradientcell{5.141}{0}{50}{white}{violet}{30} & \gradientcell{9.307}{0}{50}{white}{violet}{30} & \gradientcell{5.570}{0}{50}{white}{violet}{30} & \gradientcell{7.95}{0}{50}{white}{violet}{30} & \gradientcell{56.7330}{0}{75}{white}{violet}{30}& \gradientcell{56.6740}{0}{75}{white}{violet}{30} & \gradientcell{56.70}{0}{75}{white}{violet}{30} & \gradientcell{3.275}{0}{5}{violet}{white}{30} & \gradientcell{3.048}{0}{5}{violet}{white}{30} & \gradientcell{3.16}{0}{5}{violet}{white}{30}   \\
    2 & kNN-VC + len variation & \gradientcell{11.192}{0}{50}{white}{violet}{30} & \gradientcell{5.125}{0}{50}{white}{violet}{30} & \gradientcell{10.218}{0}{50}{white}{violet}{30} & \gradientcell{5.793}{0}{50}{white}{violet}{30} & \gradientcell{8.08}{0}{50}{white}{violet}{30} & \gradientcell{56.9488}{0}{75}{white}{violet}{30}& \gradientcell{55.638}{0}{75}{white}{violet}{30} & \gradientcell{56.29}{0}{75}{white}{violet}{30} & \gradientcell{3.28}{0}{5}{violet}{white}{30} & \gradientcell{3.387}{0}{5}{violet}{white}{30} & \gradientcell{3.33}{0}{5}{violet}{white}{30}     \\
    3 & kNN-VC+ len var + noise-in & \gradientcell{24.681}{0}{50}{white}{violet}{30} & \gradientcell{18.624}{0}{50}{white}{violet}{30} & \gradientcell{19.891}{0}{50}{white}{violet}{30} & \gradientcell{19.115}{0}{50}{white}{violet}{30} & \gradientcell{20.58}{0}{50}{white}{violet}{30} & \gradientcell{44.1260}{0}{75}{white}{violet}{30}& \gradientcell{42.3846}{0}{75}{white}{violet}{30} & \gradientcell{43.26}{0}{75}{white}{violet}{30} & \gradientcell{11.993}{0}{5}{violet}{white}{30} & \gradientcell{10.008}{0}{5}{violet}{white}{30} & \gradientcell{11.00}{0}{5}{violet}{white}{30}     \\
    4$^{\star}$ & whisper-VITS & \gradientcell{47.584}{0}{50}{white}{violet}{30} & \gradientcell{49.233}{0}{50}{white}{violet}{30} & \gradientcell{47.445}{0}{50}{white}{violet}{30} & \gradientcell{48.750}{0}{50}{white}{violet}{30} &\gradientcell{48.25}{0}{50}{white}{violet}{30} & \gradientcell{30.1074}{0}{75}{white}{violet}{30}& \gradientcell{30.5932}{0}{75}{white}{violet}{30} & \gradientcell{30.35}{0}{75}{white}{violet}{30} & \gradientcell{3.743}{0}{5}{violet}{white}{30} & \gradientcell{3.755}{0}{5}{violet}{white}{30} & \gradientcell{3.75}{0}{5}{violet}{white}{30} \\
    1 + 4$^{\star}$ & Admixture ($p=0.2$) & \gradientcell{26.003}{0}{50}{white}{violet}{30} & \gradientcell{16.155}{0}{50}{white}{violet}{30} & \gradientcell{20.776}{0}{50}{white}{violet}{30} & \gradientcell{24.722}{0}{50}{white}{violet}{30} & \gradientcell{21.91}{0}{50}{white}{violet}{30} & \gradientcell{51.2840}{0}{75}{white}{violet}{30}& \gradientcell{52.1324}{0}{75}{white}{violet}{30} & \gradientcell{51.71}{0}{75}{white}{violet}{30} & \gradientcell{3.300}{0}{5}{violet}{white}{30} & \gradientcell{3.290}{0}{5}{violet}{white}{30}  & \gradientcell{3.31}{0}{5}{violet}{white}{30} \\
    1 + 4$^{\star}$ & Admixture ($p=0.325$) & \gradientcell{34.518}{0}{50}{white}{violet}{30} & \gradientcell{32.918}{0}{50}{white}{violet}{30} & \gradientcell{34.532}{0}{50}{white}{violet}{30} & \gradientcell{33.676}{0}{50}{white}{violet}{30} & \gradientcell{33.91}{0}{50}{white}{violet}{30} & \gradientcell{49.3398}{0}{75}{white}{violet}{30}& \gradientcell{48.7304}{0}{75}{white}{violet}{30} & \gradientcell{49.04}{0}{75}{white}{violet}{30} & \gradientcell{3.514}{0}{5}{violet}{white}{30} & \gradientcell{3.336}{0}{5}{violet}{white}{30}  & \gradientcell{3.43}{0}{5}{violet}{white}{30} \\
    1 + 4$^{\star}$ & Admixture ($p=0.4$) & \gradientcell{41.192}{0}{50}{white}{violet}{30} & \gradientcell{40.660}{0}{50}{white}{violet}{30} & \gradientcell{42.182}{0}{50}{white}{violet}{30} & \gradientcell{39.225}{0}{50}{white}{violet}{30} & \gradientcell{40.81}{0}{50}{white}{violet}{30} & \gradientcell{47.0784}{0}{75}{white}{violet}{30}& \gradientcell{47.1046}{0}{75}{white}{violet}{30} & \gradientcell{47.09}{0}{75}{white}{violet}{30} & \gradientcell{3.454}{0}{5}{violet}{white}{30} & \gradientcell{3.199}{0}{5}{violet}{white}{30}  & \gradientcell{3.33}{0}{5}{violet}{white}{30} \\
    5 & WavLM Conv (base) & \gradientcell{13.622}{0}{50}{white}{violet}{30} & \gradientcell{6.987}{0}{50}{white}{violet}{30} & \gradientcell{9.307}{0}{50}{white}{violet}{30} & \gradientcell{4.231}{0}{50}{white}{violet}{30} & \gradientcell{8.54}{0}{50}{white}{violet}{30} & \gradientcell{55.5458}{0}{75}{white}{violet}{30}& \gradientcell{53.9522}{0}{75}{white}{violet}{30} & \gradientcell{54.75}{0}{75}{white}{violet}{30} & \gradientcell{3.044}{0}{5}{violet}{white}{30} & \gradientcell{2.982}{0}{5}{violet}{white}{30}  & \gradientcell{3.01}{0}{5}{violet}{white}{30} \\
    6 & WavLM Conv + Adv Spk Loss & \gradientcell{17.472}{0}{50}{white}{violet}{30} & \gradientcell{9.005}{0}{50}{white}{violet}{30} & \gradientcell{12.773}{0}{50}{white}{violet}{30} & \gradientcell{7.164}{0}{50}{white}{violet}{30} & \gradientcell{11.60}{0}{50}{white}{violet}{30} & \gradientcell{50.7706}{0}{75}{white}{violet}{30}& \gradientcell{50.4628}{0}{75}{white}{violet}{30} & \gradientcell{50.62}{0}{75}{white}{violet}{30} & \gradientcell{4.442}{0}{5}{violet}{white}{30} & \gradientcell{4.015}{0}{5}{violet}{white}{30}  & \gradientcell{4.23}{0}{5}{violet}{white}{30} \\
    7 & WavLM Conv + Discrete Loss & \gradientcell{18.041}{0}{50}{white}{violet}{30} & \gradientcell{12.268}{0}{50}{white}{violet}{30} & \gradientcell{13.716}{0}{50}{white}{violet}{30} & \gradientcell{10.913}{0}{50}{white}{violet}{30} & \gradientcell{13.73}{0}{50}{white}{violet}{30} & \gradientcell{44.5292}{0}{75}{white}{violet}{30}& \gradientcell{42.5980}{0}{75}{white}{violet}{30} & \gradientcell{43.56}{0}{75}{white}{violet}{30} & \gradientcell{10.313}{0}{5}{violet}{white}{30} & \gradientcell{10.014}{0}{5}{violet}{white}{30}  & \gradientcell{10.16}{0}{5}{violet}{white}{30} \\
    8 & WavLM Conv + Adv + Discrete Loss & \gradientcell{19.308}{0}{50}{white}{violet}{30} & \gradientcell{11.645}{0}{50}{white}{violet}{30} & \gradientcell{13.870}{0}{50}{white}{violet}{30} & \gradientcell{10.690}{0}{50}{white}{violet}{30} & \gradientcell{13.88}{0}{50}{white}{violet}{30} & \gradientcell{44.0936}{0}{75}{white}{violet}{30}& \gradientcell{42.9102}{0}{75}{white}{violet}{30} & \gradientcell{43.50}{0}{75}{white}{violet}{30} & \gradientcell{10.811}{0}{5}{violet}{white}{30} & \gradientcell{10.850}{0}{5}{violet}{white}{30}  & \gradientcell{10.83}{0}{5}{violet}{white}{30} \\
    \bottomrule
    \multicolumn{10}{l}{$^{\star}$ marks submitted systems}
  \end{tabular}
  }
\end{table*}

\subsubsection{Utility}
Two metrics are used to assess the preservation of specific acoustic attributes from the original speech. Word error rate (WER) is adopted to evaluate the ability of an anonymization system to preserve linguistic content. For emotion state preservation, unweighted average recall (UAR) is used for evaluation. Utility performances are evaluated using a pre-trained ASR system and a speech emotion recognition (SER) system, respectively. A lower WER denotes better preservation of linguistic content, while a higher UAR indicates better preservation of emotion states.

\begin{figure*}[!ht]
  \centering
  \includegraphics[width=0.9\textwidth]{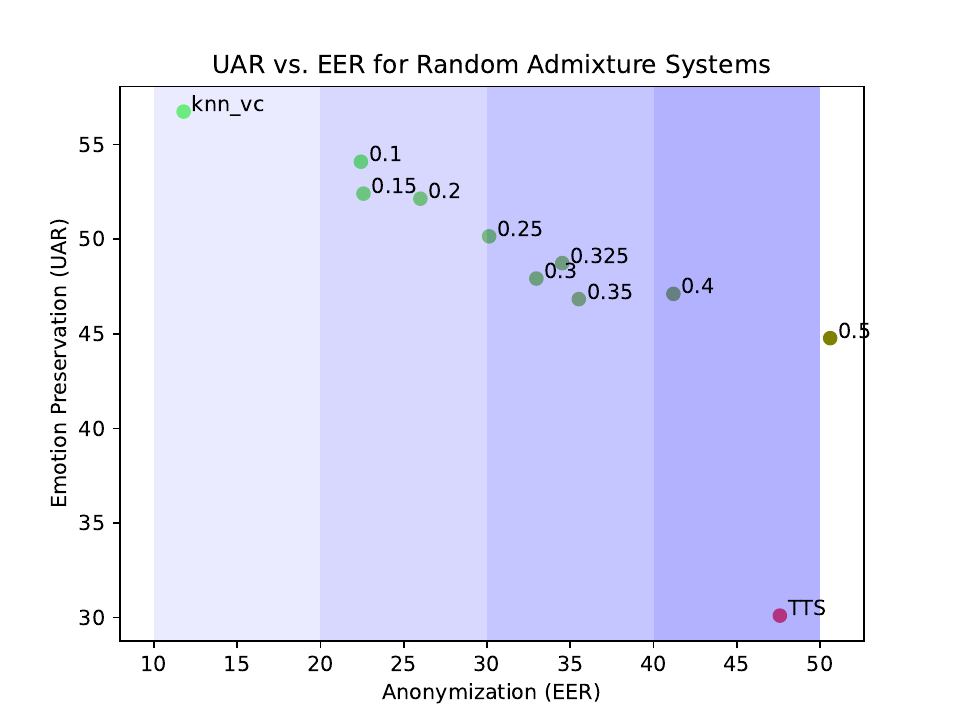}
  \caption{Scatter plot of various results from the Random Admixture system. Results from the two source systems, cascaded TTS and kNN-VC, are included. Each point is labeled and color-coded with the percentage of the admixture which was drawn from the cascaded TTS system. }
  \label{fig:results}
\end{figure*}

\subsection{Results}
The anonymization results of our systems are shown in Table \ref{tab:results}. As indicated in the table, the kNN-VC system achieves an average EER of 7.95\%, which suggests a significant exposure of the original speaker’s traits. Nevertheless, the kNN-VC system performs well in preserving emotional state and linguistic content, with an average UAR of 56.7\% and an average WER of 3.16\%, respectively. The model with speed perturbation achieves similar scores to the vanilla kNN-VC system: we found that applying a random factor between 0.8 and 1.2 to the original audio gives us marginal EER improvements without hurting UAR. Applying additive random noise drawn from a uniform distribution of vectors with infinity norm less than 32 on each of the input WavLM feature frames prior to applying kNN-VC system led to greatly improved EER at the expense of reduced UAR and WER.

The improvements we made on top of the kNN-VC system, including the neural WavLM feature conversion system, the speaker adversarial training technique, and the discrete loss, were all shown to moderately improve the anonymization performance of the system, with our WavLM Conversion system combining all features achieving a average EER of 13.88\%. However, these improvement come at the cost of greatly reducing the systems' performance on utility metrics.

On the other hand, the cascaded anonymization approach, whisper-VITS, achieves an average EER of 48.25\%, which is close to 50\%, demonstrating its ability to conceal the original speaker’s identity. Additionally, the WER utility score is 3.75\%, which is slightly lower than that of the VC approaches. Note that the WER of the transcribed text by the whisper model is 3.38\% on the libri-dev-asr set and 3.29\% on the libri-test-asr set. However, the emotion preservation performance of this approach is poor, with the system achieving an average UAR of 30.35\% on the IEMOCAP evaluation sets.

By randomly mixing the kNN-VC system with the cascaded anonymization system during voice anonymization, we were able to produce systems whose performance sit in between those of the two systems (as shown in Figure ~\ref{fig:results}), satisfying different needs for anonymization versus emotion preservation. At 50\% admixture, the EER of the system approaches 50\% while the UAR still stands at a respectable 44.77\%, stronger than any existing TTS-based anonymization system.

\section{Limitations and Future Work}

While our Random Admixture system achieved strong results in the context of the Voice Privacy Challenge setup, where the adversary uses a single model trained on anonymized data, we note that an informed adversary can effectively mitigate the impact of data-poisoning attacks at the cost of some performance degradation \cite{defend_poisoning}. We call for future studies to identify any possible equilibrium between the anonymizer and the adversary in light of admixture systems, as well as more generally, the potential role of adversarial attacks within the context of voice privacy.

We note that our TTS system does not support any type of controlled generation, which severely limits its capacity for preserving para-linguistic features such as emotion. We would like to explore investigate the effectiveness of that line of methods in our future work.

\section{Conclusion}

We experimented with two main approaches to voice anonymization: voice conversion and cascaded ASR-TTS. Our voice conversion systems generally performed strongly in emotion preservation, while our cascaded anonymization systems excel at anonymization. We were able to freely adjust the trade-off between emotion preservation and anonymization by performing a random admixture of these two systems. We call for future work to investigate if it would be possible to better the UAR-EER tradeoff curve achieved by random admixture. 

\section{Acknowledgements}

This work was supported by the Office of the Director of National Intelligence (ODNI), Intelligence Advanced Research Projects Activity (IARPA), via the ARTS Program under contract D2023-2308110001. The views and conclusions contained herein are those of the authors and should not be interpreted as necessarily representing the official policies, either expressed or implied, of ODNI, IARPA, or the U.S. Government. The U.S. Government is authorized to reproduce and distribute reprints for governmental purposes notwithstanding any copyright annotation therein.

\bibliographystyle{IEEEtran}
\bibliography{mybib}

\end{document}